\documentclass{article}
\usepackage{spconf,amsmath,graphicx}

\usepackage{amssymb}
\usepackage{stfloats}
\usepackage{amsfonts}
\usepackage{url}
\usepackage{flushend}
\usepackage{multirow}

\title{Watermarking Graph Neural Networks by Random Graphs}
\name{Xiangyu Zhao, Hanzhou Wu and Xinpeng Zhang
\thanks{It was partly supported by National Natural Science Foundation of China (Grant Nos. 61902235, U1636206, U1936214, 61525203) and also supported in part by the Shanghai ``Chenguang'' project (Grant No. 19CG46).} 
\thanks{Corresponding author: Hanzhou Wu (e-mail: h.wu.phd@ieee.org)}}
\address{Shanghai University, Shanghai 200444, China}
\begin{document}
%
\maketitle
\begin{abstract}
Many learning tasks require us to deal with graph data which contains rich relational information among elements, leading increasing graph neural network (GNN) models to be deployed in industrial products for improving service quality. However, they also raise challenges to model authentication. It is necessary to protect the ownership of the GNN models, which motivates us to watermark GNN models. In this work, an Erdos-Renyi (ER) random graph with random node feature vectors and labels is randomly generated as a trigger to train the GNN to be protected together with the normal samples. During model training, the secret watermark is embedded into the label predictions of graph nodes. During model verification, by activating a marked GNN with the trigger ER graph, the watermark can be reconstructed from the output to verify the ownership. Since the ER graph was randomly generated, by feeding it to a non-marked GNN, the label predictions of graph nodes are random, resulting in a low false alarm rate (of proposed work). Experimental results have also shown that, the performance of a marked GNN on its original task will not be impaired. And, it is robust against model compression and fine-tuning, which has shown superiority and applicability.
\end{abstract}
\begin{keywords}
Watermarking, deep learning, graph neural networks, random graph.
\end{keywords}
\section{Introduction}
Graph data has been widely used to represent the real-world relationships such as social networks. Similar to the success of deep learning (DL) models on image recognition and natural language processing, deep graph models such as graph neural networks (GNNs) \cite{paper:gcn}, \cite{paper:gat}, \cite{paper:graphSAGE}, have also achieved promising performance in processing graph data. However, training a good GNN model often requires a lot of resources such as large amounts of training data, efforts of designing and fine-tuning the model. It is important to protect the GNN model against intellectual property (IP) infringement.

As an important branch of information hiding, digital watermarking allows us to protect the IP of a digital object by slightly altering the object content for carrying the ownership information without significantly impairing the object. It is straightforward to use watermarking for protecting the IP of GNN models. However, most traditional watermarking methods are originally designed for media objects \cite{hzwu:tcsvt2017}, \cite{hzwu:wifs2016}, \cite{hzwu:ih2016}, which may be not suitable for GNN models since designing a watermarking system has to consider the statistical or structural characteristics of the object.

Actually, there are increasing watermarking methods that are designed for deep neural networks (DNNs) in recent years [7-10]. These works can be roughly organized into two categories: \emph{white box setting} and \emph{black box setting}. The former assumes that the internal details (such as network parameters and structure) of the DNN model is accessible when to extract the predefined watermark such as \cite{hzwu:mwsf2020}, \cite{Uchida2017Paper}. On the contrary, the latter is based on the scenario that ``stolen'' model is deployed as a cloud service, so only the outputted predictions of the DNN model can be acquired in the extraction phase \cite{Adi2018Paper}. Under white box setting, the watermark can be embedded into the network parameters or hidden-layer activations of a DNN. White box methods can easily be generalized to different types of DNNs but the assumption of fully accessibility to a model may be intractable in practical applications. On the other hand, watermarks used in black box methods are embedded by using some specific samples as keys to train the network so that the watermark information are carried by the prediction of these samples. However, most present black box methods are conducted on convolutional neural networks for image classification tasks, which cannot be generalized to GNNs because these two kinds of models have entirely different forms of input and output. It is therefore necessary to design a watermarking method working specifically for GNNs.

Based on the above analysis, in this paper, we present a black box watermarking method suited to GNNs. The proposed method generates an Erdos-Renyi (ER) random graph \cite{ERGraph} with random node feature vectors and labels to train the host GNN to be protected together with normal samples. During the phase of model training, the watermark is embedded into the predictions of the graph nodes. During the phase of model verification, by activating a marked GNN with the ER graph, the watermark can be identified from the output to verify the ownership. Since the ER graph was randomly generated, by feeding it to a non-marked GNN, the label predictions of the graph nodes are random, resulting in a low false alarm rate. Experimental results have demonstrated that, the performance of the marked GNN on its original task can be kept. And, it is robust against model compression and fine-tuning.

The rest is organized as follows. We first present preliminary concepts about GNNs in Section 2. Then, we detail the proposed method in Section 3, followed by experimental results Section 4. We conclude this paper in Section 5.

\begin{figure}[!t]
  \centering
  \includegraphics[width=3.5in]{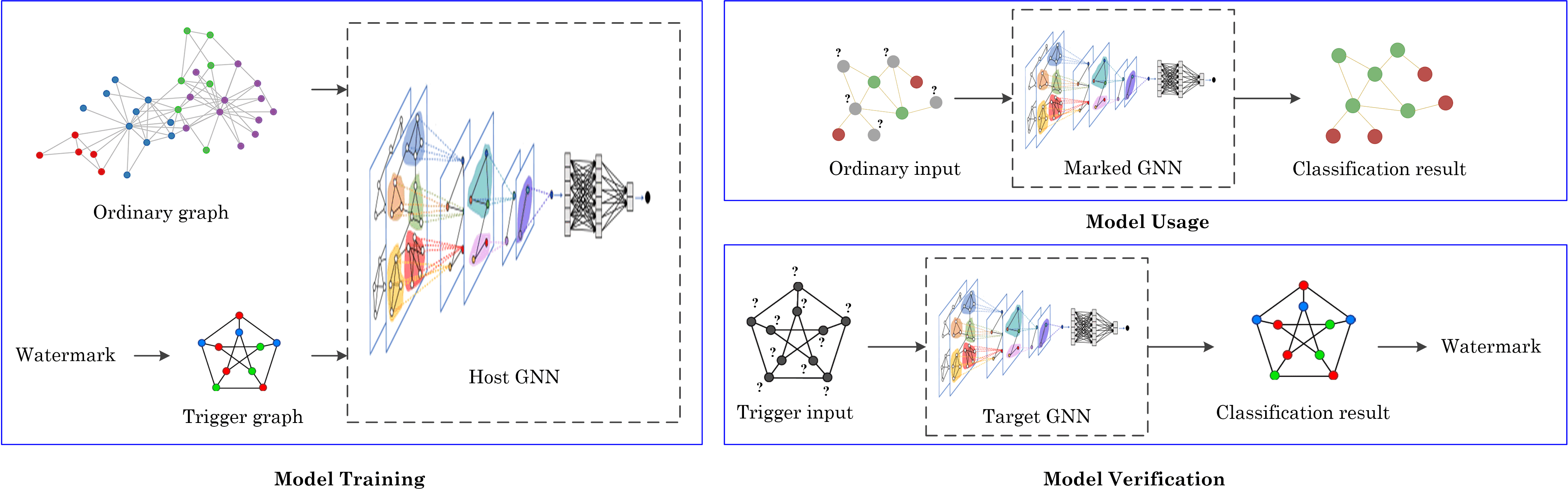}\\
  \caption{Sketch for the proposed framework suitable for node classification based GNNs (ignoring the structural details).}\label{figure1}
\end{figure}

\section{Graph Neural Networks}
Let $G = (A, X)$ be an undirected attributed graph, where $A \in \{0, 1\}^{N\times N}$ represents the adjacency matrix and $X\in R^{N\times D}$ denotes the feature matrix of nodes. Here, $N$ indicates the total number of nodes in $G$, and $D$ represents the dimension of the feature vector associated to each node. Thus, the node set of $G$ can be written as $V = \{v_1, v_2, ..., v_N\}$. Meanwhile, the edge set of $G$ can be written as $E = \{e_1, e_2, ..., e_M\}$, where $M$ means the number of edges and $e_i$ connects two nodes $v_j$ and $v_k$ in $V$, i.e., $e_i = (v_j, v_k)$. In addition, $A(a,b) = 1$ if $(v_a, v_b) \in E$, and $A(a,b) = 0$ otherwise. It is inferred that, $\sum_{i=1}^N\sum_{j=i+1}^NA(i,j) = M$. Notice that, $(v_a, v_b) \equiv (v_b, v_a)$.

The task of node classification for a graph can be formulated as follows. Given a part of labeled nodes $V_L \subsetneq V$ with labels from $Y = \{y_1, y_2, ..., y_C\}$, the target is to learn a function $f: V \rightarrow Y$ such that the labels of remaining unlabeled nodes in the graph can be accurately predicted. The majority of the existing GNNs follow a message-passing architecture: a specified node's feature representation is timely updated iteratively by aggregating information of its neighbors and its own feature, which can be formulated as:
\begin{equation}
h_{v_i}^{(t+1)} = \text{UPD}(h_{v_i}^{(t)}, \text{AGGR}(\{h_{v_j}^{(t)}|v_j\in NE(v_i)\})),
\end{equation}
where $h_{v_i}^{(t)}$ represents the feature of the node $v_i$ in the $t$-th iteration, and $NE(v_i)$ represents a set containing all neighbors of node $v_i$. Here, $\text{UPD}(\cdot)$ and $\text{AGGR}(\cdot)$ are parameterized functions with trainable parameters. These parameters are learned by optimizing the loss function indicating the difference between the ground truth labels and the prediction values.

In statistical machine learning, in the setting of \emph{transductive learning} \cite{transduction:paper}, we are given a labeled training dataset and an unlabeled testing dataset. The target of transduction is to predict the labels only for the testing samples. It is different from \emph{inductive learning} \cite{semi-supervised-learning:book}, where the goal is to find a prediction function defined on the entire space, rather than only the given testing space. When the graph is specified in advance for both model training and testing, e.g., the testing nodes in a graph are used for updating the feature representation of the training nodes in the same graph during model training, it actually corresponds to transductive learning. Graph convolutional network (GCN) \cite{paper:gcn} has shown superior performance in tackling problems in such scenario. On the contrary, if the testing samples are unseen during model training, which means the trained model will make inference on nodes dynamically inserted into the original graph or even nodes in a new graph, such circumstance actually corresponds to inductive learning. Various spatial-based approaches \cite{paper:gat}, \cite{paper:graphSAGE}, \cite{paper:powerful} are proposed to realize effective inductive learning.

Our main interest is inductive learning because inductive models are more applicable in practice, that is, a pre-trained inductive GNN model can deal with unfixed graph structure on the inference stage. Assuming that, someone is deploying his own GNN model as a cloud service, it is more likely that the deployed GNN allows customers to upload their own graph data. By contrast, transductive models can only deal with the fixed graph that appeared in the training phase, which would limit the commercial value as a DL cloud service.

\begin{figure}[!t]
  \centering
  \includegraphics[width=3.2in]{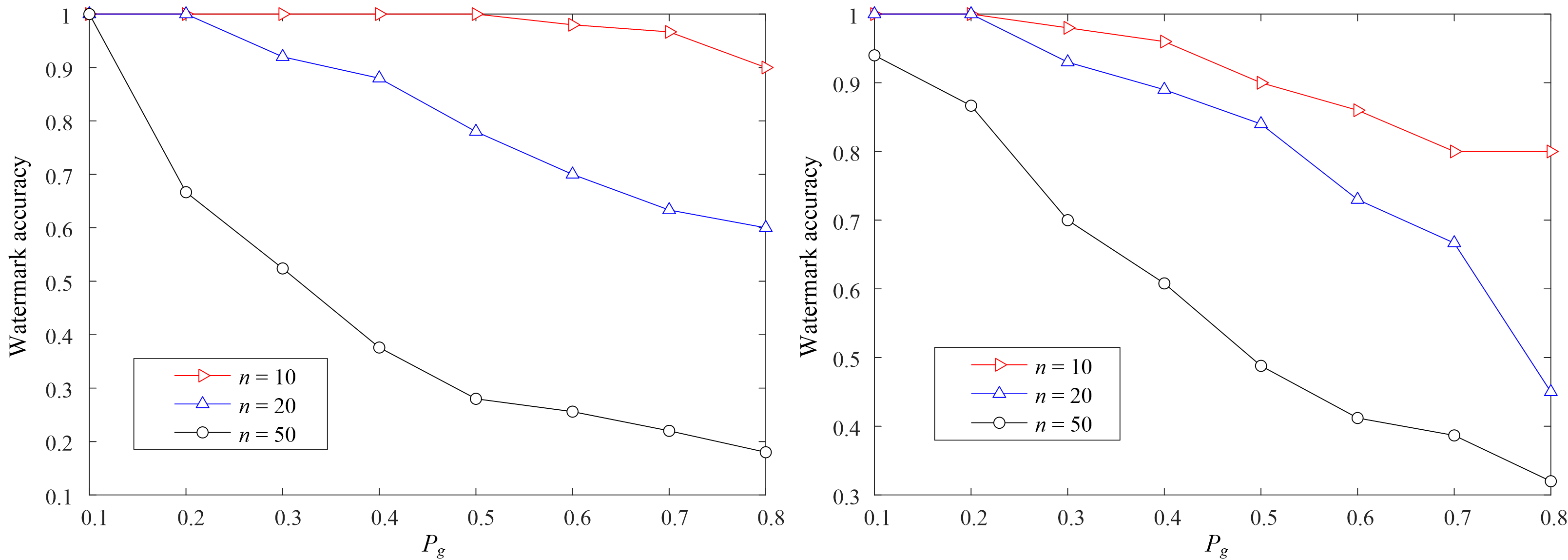}\\
  \caption{The watermark accuracy for Cora (left) and PubMed (right) due to different $P_g$. Here, $P_r = 0.1$.}\label{figure2}
\end{figure}

\begin{figure}[!t]
  \centering
  \includegraphics[width=3.2in]{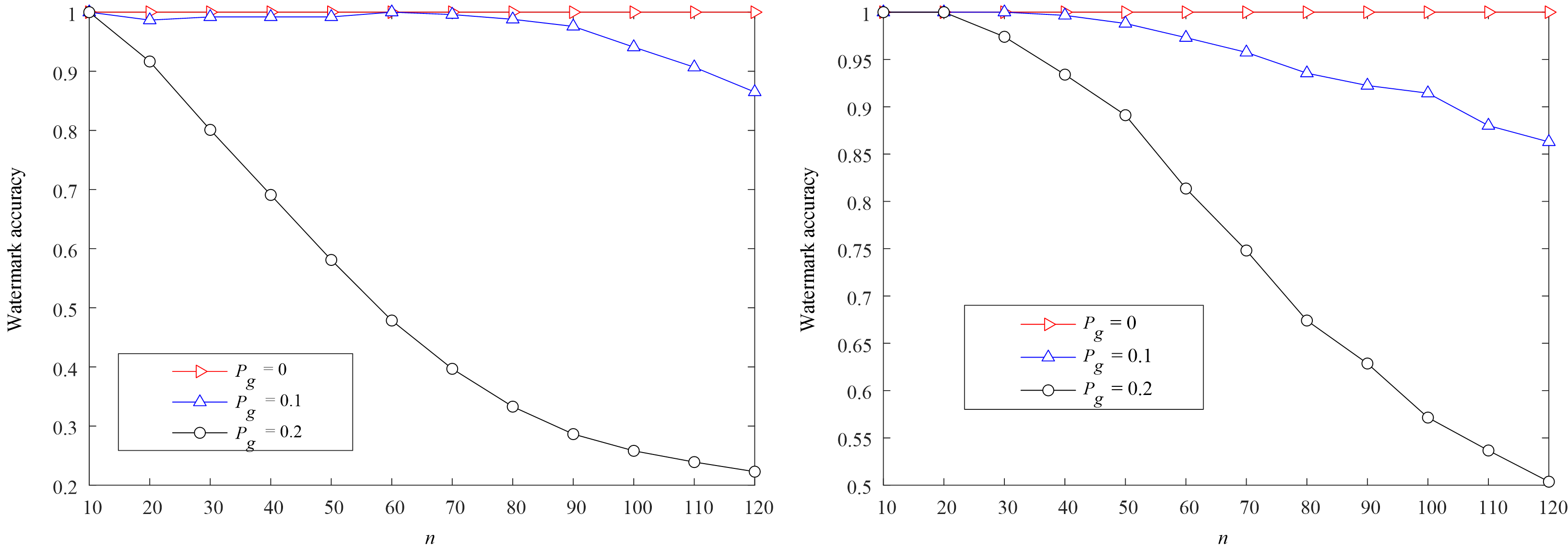}\\
  \caption{The watermark accuracy for Cora (left) and PubMed (right) due to different $n$. Here, $P_r = 0.1$.}\label{figure3}
\end{figure}

\section{Proposed Method}
In this work, we study classification based GNN. Figure 1 has shown the proposed watermarking framework, which consists of three phases, i.e., model training, model usage, and model verification. The target of model training is to train such a GNN that it can accomplish the original task while embedding a watermark into the classification result through feeding a trigger graph. Once the trained GNN model is put into use, any authenticated user can feed his/her own graph to the GNN for node classification. During model verification, the owner will feed the trigger graph into the target GNN and retrieve the watermark from the classification result for the trigger graph.

\subsection{Host Network}
We use the representative GNN named \emph{GraphSAGE (Graph SAmple and aggreGatE)} \cite{paper:graphSAGE} as the host model. GraphSAGE is an inductive node classification GNN that has been shown to be very effective in dealing with graphs with comparably low computational complexity. In GraphSAGE, to reduce the computational cost, the feature aggregator uses a fixed-size of neighbors by randomly sampling, rather than using full set of local neighbors. In \cite{paper:graphSAGE}, three aggregate functions are introduced, i.e., \emph{mean aggregator}, \emph{LSTM aggregator}, and \emph{pooling aggregator}. In this paper, we take the mean aggregator as the benchmark due to its simplicity and efficiency. GraphSAGE with the mean aggregator (GraphSAGE-mean) can be regarded as an inductive version of GCN \cite{paper:gcn}. Specifically, an iterative layer of GraphSAGE-mean can be expressed as:
\begin{equation}
h_{v_i}^{(t+1)} = \sigma(W^{(t+1)}\cdot \text{M}(\{h_{v_i}^{(t)}\}\cup \{h_{v_j}^{(t)}|v_j\in NE(v_i)\})),
\end{equation}
where $\sigma(\cdot)$ represents the activation function, $W^{(t+1)}$ is a trainable matrix and $\text{M}(\cdot)$ shows the element-wise mean operator. We refer a reader to \cite{paper:graphSAGE} for more details.

\begin{table*}[!t]
\centering
\caption{Node-classification accuracy due to different parameter settings. (Baseline is non-marked.)}
\renewcommand{\arraystretch}{1}
	\scalebox{0.8}{
	\begin{tabular}{|c|c|c|c|c|c|c|c|c|c|c|c|}
		\multicolumn{1}{c}{}\\
		\hline
		\multirow{3}{*}{Dataset} & \multirow{3}{*}{
			\begin{tabular}{c}
				Baseline
			\end{tabular}
		}	
		& \multirow{3}{*}{$n$}
		& \multicolumn{9}{c|}{Node-classification accuracy (marked)} \\ \cline{4-12}
		& & & \multicolumn{3}{c|}{$P_r=0.1$}
		& \multicolumn{3}{c|}{$P_r=0.2$}
		& \multicolumn{3}{c|}{$P_r=0.3$} \\ \cline{4-12}
		& & & $P_g=0$ & $P_g=0.1$ & $P_g=0.2$
			& $P_g=0$ & $P_g=0.1$ & $P_g=0.2$
			& $P_g=0$ & $P_g=0.1$ & $P_g=0.2$  \\
		\hline
		\multirow{3}{*}{Cora} & \multirow{3}{*}{0.840} & 10
		& 0.828 & 0.828 & 0.829 & 0.829 & 0.829 & 0.831 & 0.832 & 0.829 & 0.828     \\ \cline{3-12}
		& & 50 & 0.829 & 0.828  & 0.826 & 0.831 & 0.830 & 0.827 & 0.829 & 0.842 & 0.841 \\ \cline{3-12}
		& & 100 & 0.832 & 0.831 & 0.825 & 0.829 & 0.843 & 0.842 & 0.834 & 0.842 & 0.838 \\
		\hline
		\multirow{3}{*}{PubMed} & \multirow{3}{*}{0.862} & 10 & 0.843 & 0.842 & 0.843 & 0.847 & 0.841 & 0.844 & 0.848 & 0.847  & 0.849 \\
		\cline{3-12}
		& & 50 & 0.839 & 0.842 & 0.841 & 0.840 & 0.844 & 0.841 & 0.847 & 0.848 & 0.843 \\
		\cline{3-12}
		& & 100 & 0.841 & 0.843 & 0.846 & 0.842 & 0.843 & 0.848 & 0.846 & 0.844 & 0.851 \\
		\hline
	\end{tabular}}
\label{tab_fidelity}
\end{table*}

\begin{table*}[!t]
    \centering
	\caption{Experimental results on evaluating the watermark uniqueness.}
	\renewcommand{\arraystretch}{1}
	\scalebox{0.8}{
		\begin{tabular}{|c|c|c|c|c|c|c|c|c|c|c|}
			\multicolumn{1}{c}{}\\
			\hline
			\multirow{3}{*}{Dataset}
			& \multirow{3}{*}{$n$}
			& \multicolumn{9}{c|}{Watermark accuracy (non-marked)} \\ \cline{3-11}
			& & \multicolumn{3}{c|}{$P_r=0.1$}
			& \multicolumn{3}{c|}{$P_r=0.2$}
			& \multicolumn{3}{c|}{$P_r=0.3$} \\ \cline{3-11}
			& & $P_g=0$ & $P_g=0.1$ & $P_g=0.2$
			& $P_g=0$ & $P_g=0.1$ & $P_g=0.2$
			& $P_g=0$ & $P_g=0.1$ & $P_g=0.2$  \\
			\hline
			\multirow{3}{*}{Cora} & 10
			& 0.18 & 0.11 & 0.11 & 0.11 & 0.13 & 0.12 & 0.14 & 0.11 & 0.08     \\ \cline{2-11}
			& 50 & 0.15 & 0.14  & 0.13 & 0.15 & 0.14 & 0.10 & 0.16 & 0.14 & 0.10 \\ \cline{2-11}
			& 100 & 0.13 & 0.14 & 0.16 & 0.14 & 0.17 & 0.16 & 0.13 & 0.16 & 0.16 \\
			\hline
			\multirow{3}{*}{PubMed} & 10 & 0.30 & 0.35 & 0.33 & 0.34 & 0.33 & 0.28 & 0.33 & 0.32  & 0.35 \\
			\cline{2-11}
			& 50 & 0.35 & 0.34 & 0.35 & 0.33 & 0.34 & 0.33 & 0.33 & 0.34 & 0.35 \\
			\cline{2-11}
			& 100 & 0.34 & 0.33 & 0.30 & 0.33 & 0.34 & 0.33 & 0.32 & 0.34 & 0.33 \\
			\hline
	\end{tabular}}
	\label{tab_intergrity}
\end{table*}

\subsection{Trigger Graph}
Random graphs involve graph theory and probability theory. A random graph including $n$ nodes and $m$ edges is generally generated by starting with $n$ isolated nodes and inserting $m$ edges between nodes at random. Different random graph generators show different probability distributions over graphs. A classical model is presented in \cite{ERGraph}, in which each possible edge occurs independently with a probability $p \in (0, 1)$. It is also called Erdos-Renyi (ER) random graph model. In the model, the probability of obtaining any one particular random graph with $n$ nodes and $m$ edges is $p^m(1-p)^{n(n-1)/2-m}$. In this paper, we will use an ER random graph as the trigger input for watermark embedding and verification. We refer a reader to \cite{ERGraph} for more details about ER graph.

\begin{figure}[!t]
  \centering
  \includegraphics[width=3.2in]{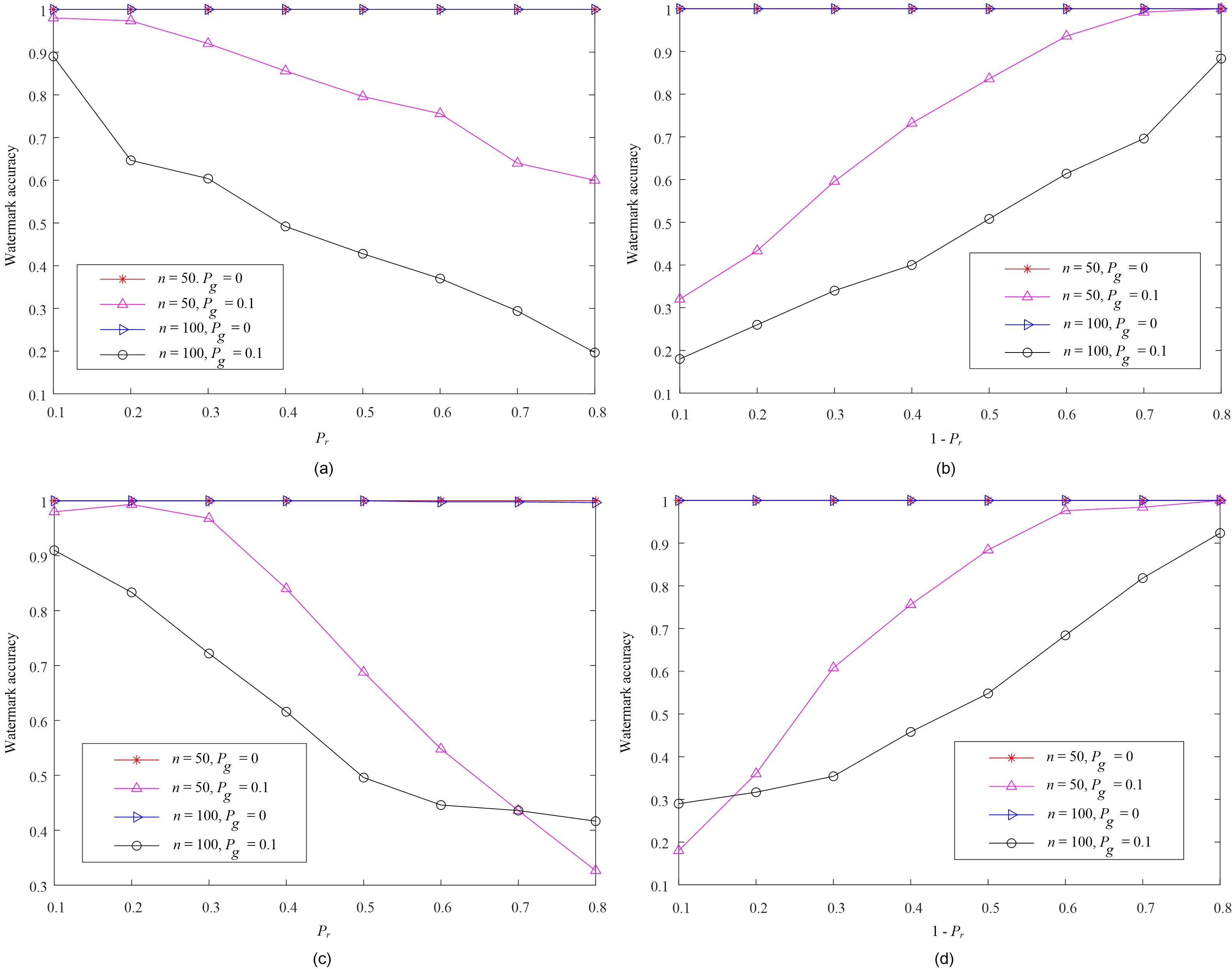}\\
  \caption{Watermark accuracy for Cora (a, b) and PubMed (c, d) due to different $P_r$. Here, (b) is obtained by flipping the binary feature vectors of nodes of (a), and (d) is obtained by flipping the binary feature vectors of nodes of (c). The pre-trained GNNs for (a, b, c, d) are different from each other.
  }\label{figure4}
\end{figure}

\begin{figure}[!t]
  \centering
  \includegraphics[width=3.2in]{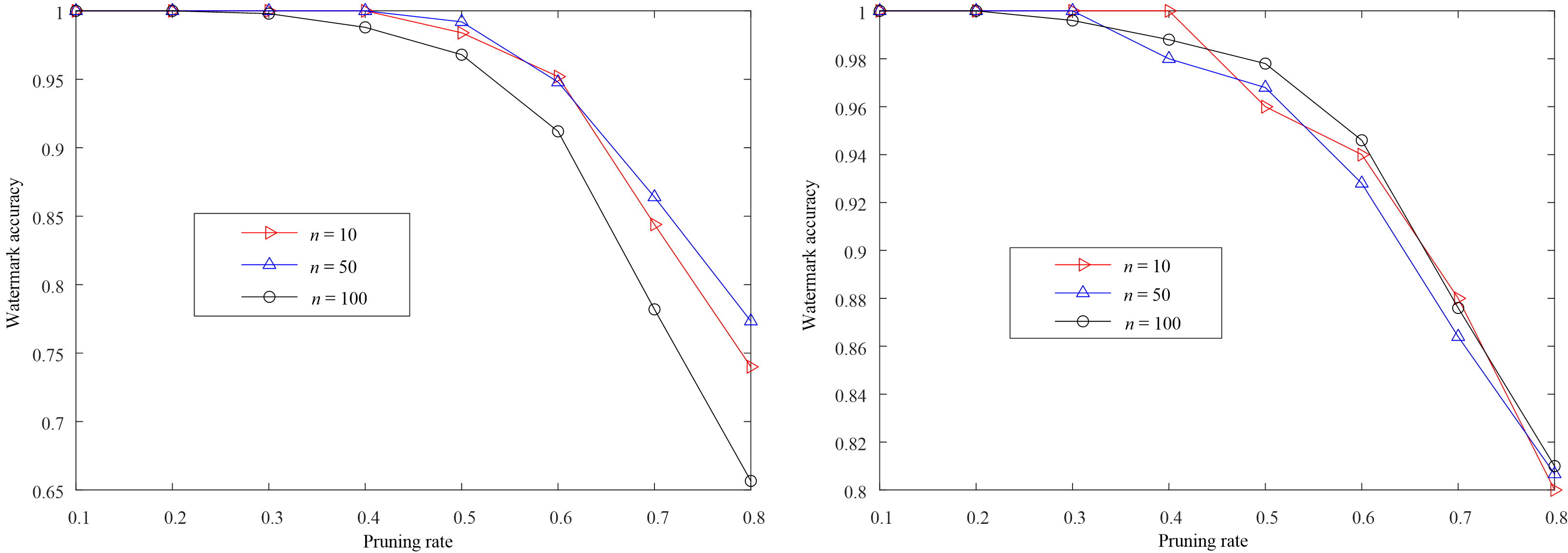}\\
  \caption{The watermark accuracy for Cora (left) and PubMed (right) due to model pruning. Here, $P_g = 0$ for all cases.
  }\label{figure5}
\end{figure}

An attributed ER random graph with node labels will be randomly generated as the trigger graph. The watermark information is carried by the labels of the trigger graph nodes. Assuming that, the value of the label of a node is an integer in range $[0, C)$, where $C \geq 2$. It can be therefore considered that the watermark is a $C$-ary sequence, i.e., $\{b_1, b_2, ..., b_n\}$, where $b_i\in [0,C)$ and $n$ means the number of nodes in the trigger graph. Obviously, given $n$, the watermark (labels) can be randomly determined in advance. For the attributed ER random graph, the dimension of the feature vector associated to each node and the predictive category of each node should be aligned with the original dataset. For compactness, we use the aforementioned $G = (A, X)$ to represent the trigger graph, from which we can infer that $N = n$ and $M = m$. The structure of $G$ is controlled by two factors, i.e., the total number of nodes $N$ and the probability of the edge existence $P_g$. Besides the generation of graph structure, the feature matrix $X$ can be filled with 0 and 1 randomly, rather than random real numbers. The rationale behind it is that the node features of many graph datasets are discrete and it is demonstrated that GNNs can effectively learn the pattern from such form of features. To this end, we first initialize $X$ as an all-zero matrix and then uniformly randomly set a proportion of the elements in the matrix to be 1. The proportion of ones is denoted as $P_r$. In this way, three parameters $N$, $P_g$ and $P_r$ would have significant impact on the generation of the ER trigger graph.

\subsection{Watermark Embedding and Extraction}
Once the trigger graph carrying a secret watermark has been generated, we can embed the secret watermark into the host GNN by training the GNN with the trigger graph. The reason is, training will force the GNN to learn the mapping between the graph nodes and the corresponding labels. It is noted that, the GNN should be also fed with normal graph data.

To verify the ownership of a target GNN, one has to reconstruct the trigger graph according to the secret key and then feed the trigger graph into the GNN model to obtain the classification results of the graph nodes. By comparing the watermark reconstructed from the classification results and the expected watermark, we can identify the ownership of the GNN model. The watermark accuracy can be defined as the percentage of correctly extracted elements. Obviously, it yields \emph{one-bit watermarking} if we relax the accuracy, i.e., the authorship is verified if the accuracy is higher than a threshold.

\begin{table}[!t]
    \centering
	\caption{Accuracy after fine-tuning. $P_g = 0, P_r = 0.1$.}
	\renewcommand{\arraystretch}{1}
	\scalebox{0.8}{
	\begin{tabular}{|c|c|c|c|}
		\multicolumn{1}{c}{}\\
		\cline{1-4}
		Dataset & $n$ & Fine-tuning epochs & Watermark accuracy  \\ \cline{1-4}
		
		\multirow{9}{*}{Cora} & \multirow{3}{*}{10} & 10 & 1.00  \\
		\cline{3-4}
		
		& & 20 & 1.00  \\
		\cline{3-4}
		
		& & 30 & 1.00  \\
		\cline{2-4}
		
		& \multirow{3}{*}{50} & 10 & 1.00  \\
		\cline{3-4}
		
		& & 20 & 0.96  \\
		\cline{3-4}
		
		& & 30 & 0.98  \\
		\cline{2-4}
		
		& \multirow{3}{*}{100} & 10 & 1.00  \\
		\cline{3-4}
		
		& & 20 & 0.99  \\
		\cline{3-4}
		
		& & 30 & 0.97  \\
		\cline{1-4}

		\multirow{9}{*}{PubMed} & \multirow{3}{*}{10} & 10 & 1.00  \\
		\cline{3-4}
		
		& & 20 & 1.00  \\
		\cline{3-4}
		
		& & 30 & 1.00  \\
		\cline{2-4}
		
		& \multirow{3}{*}{50} & 10 & 0.98  \\
		\cline{3-4}
		
		& & 20 & 0.98  \\
		\cline{3-4}
		
		& & 30 & 0.94  \\
		\cline{2-4}
		
		& \multirow{3}{*}{100} & 10 & 0.89  \\
		\cline{3-4}
		
		& & 20 & 0.88  \\
		\cline{3-4}
		
		& & 30 & 0.84  \\
		\cline{1-4}
	\end{tabular}
	}
\label{tab_modelft}
\end{table}

\section{Experimental Results and Analysis}
\subsection{Setup}
We conduct experiments on two representative datasets: Cora \cite{cora} and PubMed \cite{pubmed}. The former\footnote{\url{https://relational.fit.cvut.cz/dataset/CORA}} consists of 2708 publications classified into one of seven classes. The citation network consists of 5429 links. Each publication is described by a binary valued word vector indicating the absence/presence of the corresponding word from the dictionary having 1433 unique words. The latter\footnote{\url{https://linqs-data.soe.ucsc.edu/public/Pubmed-Diabetes.tgz}} consists of 19717 publications classified into one of three classes. The citation network consists of 44338 links. Each publication is described by a TF/IDF (term frequency-inverse document frequency) weighted word vector from a dictionary having 500 unique words. For both datasets, we use 40\% nodes for training, 20\% nodes for validation and the rest for testing. An experiment was repeated with ten times to obtain the averaged result.

We use GraphSAGE-mean, an inductive learning model, as the host GNN. the model is composed of two layers, where the first layer has 128 neurons and the second one has 128 neurons with the softmax output. For both layers, we set the sampling number as 5 and activation function as ReLU \cite{relu:paper}. We train the host GNN by minimizing the cross entropy loss with Adam optimizer \cite{adam2014Paper}, where the batch size is set to 128.

\subsection{Network Fidelity}
Network fidelity means the watermarking procedure should not significantly degrade the performance of the GNN model on its own original task. To evaluate the network fidelity, we train two GNN models, i.e., \emph{marked} GNN and \emph{non-marked} GNN. While the non-marked GNN uses the normal graph data for training, the marked version uses both the normal graph data and the trigger graph. Table 1 shows the experimental results. It can be seen from Table 1 that, though the performance will decline after embedding a watermark, the degradation can be kept with a low level, which has verified the applicability of the proposed method. Moreover, though different parameters result in difference performance, they are close to each other. It indicates that, benefiting from the strong representation ability of GNN, in terms of network fidelity, the watermark does not impair the original task.

\subsection{Watermark Reliability and Capacity}
Watermark reliability reveals the difference between the extracted watermark and the original watermark. It is necessary that the difference can be as low as possible so that the ownership can be reliably identified. Figure 2 shows the watermark accuracy for the Cora and PubMed datasets due to different parameter settings. It can be observed that, in terms of graph size, a smaller number of graph nodes leads to the higher watermark accuracy. It is reasonable as a relatively larger-scale graph requires the host GNN to have stronger representation ability. It can be also seen from Figure 2 that, for a fixed number of graph nodes, as the probability of edge existence $P_g$ increases, the watermark accuracy will decline. A reasonable explanation is that the graph structure and node features are generated randomly, meaning that, adding edges between nodes with independently random features is to correlate independent features forcibly, which introduces ``unreasonable'' random pattern for the GNN model. It means that, adding more edges between nodes (with random features) increases the difficulty of graph pattern learning, which can reduce the classification accuracy. Figure 3 has further verified the above analysis. In addition, in terms of capacity, by using suitable parameters, more watermark information can be embedded (see Figure 3) since more nodes means higher capacity.

\subsection{Feature Sensitivity}
Unlike previous arts designed for CNN models, in this work, we have to generate feature vectors for graph nodes in a random fashion. It is necessary to analyze the impact of the random features on the performance of watermark reconstruction. In the proposed method, the feature vectors are generated as random binary strings. The percentage of ``1''s in the binary strings, i.e., $P_r$, can be used to characterize the random features. Intuitively, it is guessed that when $P_r$ is close to zero or one, the watermark accuracy should be high, and; the watermark accuracy should be low when $P_r$ is close to 0.5 since $P_r = 0.5$ corresponds to the maximum entropy. However, Figure 4 (a) and (c) show the experimental results on the two datasets due to different $P_r$. It can be observed that, surprisingly, as $P_r$ increases, the performance will decline for positive $P_g$. It implies that, more ``1''s will increase the difficulty of watermark pattern learning. To further verify it, we flip the feature vectors corresponding to Figure 4 (a) and (c), and obtain the watermark accuracy shown in Figure 4 (b) and (d). It means that, ``0'' and ``1'' should not be considered as equally important in the feature vectors, which is due to the non-linear and asymmetric learning in graph models. In addition, when $P_g = 0$, the watermark in most cases can be perfectly reconstructed, showing that, an trigger graph consisting of many isolated nodes would be the best for watermark embedding.

\subsection{Watermark Uniqueness}
The watermark uniqueness means that only the marked GNN model leads us to retrieve the watermark, while any non-marked GNN model should reveal nothing about the watermark. It is expected that, the averaged watermark accuracy for non-marked GNN models should be close to $1/C$, where $C$ is the total number of label categories. Table 2 has shown the watermark accuracy of the non-marked GNN models due to different trigger graphs. As shown in Table 2, experimental results meet the above inference, verifying that the proposed work ensures watermark uniqueness.

\subsection{Model Robustness}
In applications, a marked GNN model may be altered, leading the embedded watermark to be undetectable. To this end, we have to test the robustness. In this work, we consider two common attacks, i.e., model pruning and model fine-tuning.

\subsubsection{Model Pruning}
Model pruning sets some parameters to zero while maintaining the performance on the primary task. In our experiments, to mimic pruning attack, we set a certain number of parameters with smallest absolute values to zero and test whether the hidden watermark can be still reconstructed or not. We define the ratio between the number of pruned parameters and the total number of parameters as pruning rate. It can be observed from Figure 5 that, for both datasets, the watermark can be extracted near-perfectly even 50\% of the parameters are pruned, indicating that, the proposed work can well resist against model pruning. We can also observe that, overall, a larger watermark size is less robust against model pruning, which is intuitive and reasonable since a larger trigger set requires altering more model parameters and model pruning is more likely to erase those parameters that are responsible for the pattern learning of comparably larger trigger set. Therefore, in practice, we have to make a balance between the watermark size and robustness. Notice that, in Figure 5, we have $P_g = 0$ for all cases, implying that, the watermark accuracy are all equal to 1 for non-pruned networks (refer to Figure 4).

\subsubsection{Model Fine-tuning}
Model fine-tuning is another common technique operating on a pre-trained model, which is likely to be used by an adversary to remove the watermark. In our experiments, we use 30\% of the test set along with the training set to fine-tune the marked GNN and evaluate the effect of the fine-tuning procedure on the watermark reconstruction. We constrain the number of fine-tuning epochs to be less than that in the original training phase of the host GNN because we assume that the computational resource of the adversary is restricted so s/he has to ``steal'' models from others. As shown in Table 3, for both datasets, the watermark accuracy of the fine-tuned GNN on the trigger set degrades in a very limited range as the number of fine-tuning epochs increases. When the watermark size is small (e.g., ten nodes used in the trigger set), model fine-tuning does not impair the watermark extraction at all.

\section{Conclusion}
In this paper, we present the first watermarking framework to GNNs. A random graph associated with features and labels is generated as the trigger input. By training the host GNN with the trigger graph, the watermark can be identified from its output during verification. Experiments have shown that, the proposed watermarking system shows superior performance in network fidelity, watermark reliability, watermark uniqueness as well as robustness. With the fast development of graph data processing, it will be very necessary to protect graph learning models. This work makes a preliminary attempt protecting the IP of GNNs. In future, we will explore methods that are more robust and reliable and study how to embed watermarks into various types of GNNs besides node level tasks.


\end{document}